\documentstyle[psfig,12pt]{article}

\newcommand{\lapprox}{%
\mathrel{%
\setbox0=\hbox{$<$}
\raise0.6ex\copy0\kern-\wd0
\lower0.65ex\hbox{$\sim$}
}}
\newcommand{\gapprox}{%
\mathrel{%
\setbox0=\hbox{$>$}
\raise0.6ex\copy0\kern-\wd0
\lower0.65ex\hbox{$\sim$}
}}

\begin{document}

\font\fifteen=cmbx10 at 15pt
\font\twelve=cmbx10 at 12pt

\begin{titlepage}

\begin{center}

\renewcommand{\thefootnote}{\fnsymbol{footnote}}

{\twelve Centre de Physique Th\'eorique\footnote{
Unit\'e Propre de Recherche 7061
}, CNRS Luminy, Case 907}

{\twelve F-13288 Marseille -- Cedex 9}

\vspace{1 cm}

{\fifteen COMMENT ON THE PREDICTION OF TWO-LOOP \\
STANDARD CHIRAL PERTURBATION THEORY \\
FOR LOW-ENERGY $\mbox{\Large{$\pi\pi$}}$ SCATTERING}

\vspace{0.3 cm}

\setcounter{footnote}{0}
\renewcommand{\thefootnote}{\arabic{footnote}}

{\bf
L. GIRLANDA\footnote{Division de Physique Th\'eorique, Institut de Physique Nucl\'eaire, F-91406 Orsay Cedex, France. Division de Physique Th\'eorique is
Unit\'e de Recherche des Universit\'es Paris XI et Paris VI
associ\'ee au CNRS}, M. KNECHT, B. MOUSSALLAM$^1$ and 
J. STERN$^1$
}

\vspace{2,3 cm}

{\bf Abstract}

\end{center}

Four of the six parameters defining the two-loop $\pi \pi$ scattering
amplitude have been determined using Roy dispersion relations. Combining
this information with the Standard $\chi$PT expressions, we obtain the
threshold parameters, low-energy phases and the $O(p^4)$ constants $l_1^r$,
$l_2^r$. The result $ l_2^r ( M_{\rho} ) = (1.6 \pm 0.4 \pm 0.9) \times 10^{-3}$
 (${\bar l_2} = 4.17 \pm 0.19 \pm 0.43$) reproduces the correct $D$-waves but it is incompatible with existing
Standard $\chi$PT analyses of $K_{l4}$ form factors beyond one loop.

\vspace{2 cm}


\bigskip

\noindent Number of figures: 2

\bigskip

\noindent March 1997

\noindent CPT-97/P.3470

\noindent IPNO/TH 97-08

\bigskip

\noindent anonymous ftp or gopher: cpt.univ-mrs.fr

\renewcommand{\thefootnote}{\fnsymbol{footnote}}

\end{titlepage}

\setcounter{footnote}{0}

{\bf 1}. During the last few years there has been a noticeable revival of interest in
the high precision analysis of low-energy $\pi\pi$ scattering
\cite{knecht}-\cite{FSS}. There are at
least two reasons for this. First, it has been shown
\cite{FSS,knecht,knecht2} and repeatedly
emphasized \cite{stern} that  the $\pi\pi$ scattering amplitude in the threshold region
is particularly sensitive to the strength of quark anti-quark pair
condensation in the QCD vacuum: the smaller the condensate, the stronger the
isoscalar $S$-wave $\pi \pi$ interaction. The accurate measurement of
$S$-wave scattering lengths would, indeed, provide the first experimental
evidence in favour of, or against, the standardly admitted hypothesis
according to which the mechanism of spontaneous chiral symmetry breaking is
dominated by the formation of a large  $<\bar  q q>$ condensate. Within QCD,
this hypothesis is by no means a logical necessity and its experimental test
might well become an important step towards a non-perturbative understanding
of the quark-gluon dynamics. The second reason which makes detailed $\pi 
\pi $ studies topical, is that there are two new high precision experiments currently
under preparation: $i)$ The phase shift difference $\delta_0^0 (E) -
\delta_1^1 (E) $ at low energies ($ E < 400 \mbox{MeV}$) will be extracted
from a new $K_{l4}$-decay experiment \cite{2hand} performed with the  KLOE
detector at the
Frascati $\phi$-factory DA$\Phi$NE \cite{dafne}. $ii)$ At CERN, the  project DIRAC
\cite{dirac} aims at
the measurement of the lifetime of $\pi^{+} \pi^{-}$ atoms to $10 \%$,
implying
the determination of the combination of scattering lengths $\mid a_0^0 -
a_0^2 \mid$  with a  $5 \%$
 accuracy. On the theoretical side an even better precision can be reached
 by a systematic use of chiral perturbation theory \cite{weinberg,gale} ($\chi$PT). The
 low-energy expansion of the $\pi \pi$ scattering amplitude $A(s | t,u)$ starts at
 order $O(p^2)$ given by Weinberg more than 30 years ago \cite{wein}. Subsequently, the
 one-loop $O(p^4)$ contribution to $A(s|t,u)$ has been calculated by Gasser
 and Leutwyler \cite{gale2,gale}. It is given by four low-energy constants $l_1$, $l_2$,
 $l_3$, $l_4$ besides the (charged) pion mass $M_{\pi}$ and the decay
 constant $F_{\pi}$. The present state of the art involves the two-loop $O(p^6)$
 order and the present letter concerns this degree of accuracy.
 
 \vspace{.5cm}

{\bf 2}. The $O(p^6)$ amplitude $A(s| t,u)$ has been first given in Ref.~\cite{knecht} in the
 form \begin{equation}
 A(s|t,u) = A_{KMSF}(s|t,u;\alpha,\beta;\lambda_1, \lambda_2, \lambda_3,
 \lambda_4) + O \left[ \left({p \over \Lambda_H} \right)^8 , \left( {M_{\pi} \over
 \Lambda_H} \right)^8 \right].
 \end{equation}
 The function $A_{KMSF}$, which depends on the  Mandelstam variables $s$,
 $t$, $u$ and on the six
 parameters $\alpha$, $\beta$, $\lambda_1$, \ldots $\lambda_4$, is explicitly
 displayed in \cite{knecht}. Here, $p$ denotes the characteristic pion momentum and
 $\Lambda_H$ is the mass scale of bound states not protected by the chiral
 symmetry, $\Lambda_H \sim 4 \pi F_{\pi} \sim 1 \mbox{ GeV}$. The result (1)
 holds independently of the strength of the quark condensate. The latter
 merely shows up in the size of the constant $\alpha$: for standard, large
 values\footnote{ $<\bar q q>$ denotes the single flavour condensate in the $
 SU(2) \times SU(2)$ chiral limit $m_u = m_d = 0$ at the QCD scale $\nu = 1$
 GeV.} $<\bar{q} q>  \simeq - (250 \mbox{ MeV} )^3$ one has $ \alpha \simeq
 1$ and its value increases up to $ \alpha \simeq 4$ for $ \mid <\bar{q} q> \mid$
 decreasing down to zero. The parameter $\beta$ is less sensitive to the
 value of the condensate, remaining always close to unity. It has been shown
\cite{knecht2} that the remaining four constants $\lambda_1, \ldots \lambda_4$ can be
 rather accurately determined from the existing $\pi \pi$ scattering data
 \cite{hyams} in
 the intermediate energy range $ 0.5 \mbox{ GeV} < E < 1.9 \mbox{ GeV}$ using
 the Roy dispersion relations \cite{roy}. The latter explicitly incorporate crossing
 symmetry and consequently they strongly constrain the $\pi \pi$ amplitude at
 low energies. Equating the perturbative formula (1) with the Roy dispersive
 representation in a whole low-energy region of the Mandelstam plane, one
 infers the values of $\lambda_1$, \ldots $\lambda_4$, whereas the
 paramenters $\alpha$ and $\beta$ remain essentially undetermined. The
 resulting $\lambda_i$'s are almost independent of $\alpha$ and $\beta$. Here
 we quote and use the central values corresponding to $\alpha = 1.04$,
 $\beta= 1.08$  \cite{knecht2},
 \begin{eqnarray}
&\lambda_1 = (-5.7 \pm 2.2) \times 10^{-3}, & \lambda_2 = (9.3 \pm 0.5)
 \times 10^{-3}, \nonumber \\
&\lambda_3 = (2.2 \pm 0.6) \times 10^{-4}, & \lambda_4 = (-1.5 \pm 0.12)
 \times 10^{-4}.  
\end{eqnarray}

The quoted errors include experimental uncertainties on $\pi\pi$
phase-shifts and inelasticities in the medium energy region and an estimate
of the  systematic
error arising from neglected higher orders in the low-energy representation
(1). The errors due to the uncertainty in the high-energy behaviour of the
$\pi \pi$  scattering amplitude are negligible.

\vspace{.5cm}

{\bf 3}. With the constants $\lambda_i$ determined,  Eq.~(1) allows one to convert
new high-precision experimental information on low-energy $\pi \pi$ phase
shifts and/or threshold parameters into a measurement  of $\alpha$ and
$\beta$  and finally, into an experimental determination of the quantity 
$(m_u + m_d) <\bar{q} q>$ (the detailed relation between $\alpha$ and $\beta$ and the
condensate can be found in Ref.~\cite{knecht}). Conversely, Eq.~(1) can be used to
predict, for each value of the condensate, all low-energy observables.
It is of particular importance to assess with as much accuracy as possible
the prediction concerning the standard alternative of a large
$<\bar q q>$  condensate. The strength of the $<\bar{q}
q>$  condensate is conveniently described  by the deviation from the
Gell-Mann--Oakes--Renner relation, i.e. by the parameter 
\begin{equation}
{m \over m_0} = { {F_{\pi}^2 M_{\pi}^2} \over { 2 m \mid < \bar{q} q >
\mid}} - 1\,. 
\end{equation}
Here, $ m = {1 \over 2} ( m_u + m_d)$ is the running quark mass and $m_0$ is
a mass scale characteristic of $\bar q q$ condensation. The standard
alternative of a large condensate corresponds to $m_0 \gapprox \Lambda_H$. In
this special case the ratio (3) can be treated as an expansion parameter,
$m/ m_0 = O( p^2 / \Lambda_H^2)$ and the general low-energy expansion
becomes  the standard chiral perturbation theory (S$\chi$PT)
\cite{gale}.
The complete S$\chi$PT two-loop calculation of the $\pi \pi$-scattering
amplitude has been recently completed by Bijnens {\it et al.} \cite{bijn}. Not
surprisingly, this calculation recovers the formula (1) giving, in addition,
the expressions of the six parameters $\alpha, \beta, \lambda_1, \ldots,
\lambda_4$ in terms of $i)$ $M_{\pi}$, $F_{\pi}$, $ii)$ four $O(p^4)$
constants $l_1^r(\mu)$, $l_2^r(\mu)$, $l_3^r(\mu)$ and  $l_4^r(\mu)$  and
finally $iii)$ six $O(p^6)$ constants $r_1^r(\mu), \ldots, r_6^r(\mu)$ which
appear in the effective lagrangian and are renormalized  at a scale $\mu$.
These expressions read\\
\begin{eqnarray}
\alpha &=&1 + \left( -{{1}\over 2}\, L + 6\,{l_3^r} + 2\,{l_4^r} - {1\over
{32\,{{\pi }^2}}} \right) \,{M_{\pi}^2 \over F_{\pi}^2} +  \left[ -8\,{k_1} - {{14}\over 3}\, k_2 - 13\,{k_3} -
{{3}\over 2}\, k_4 \right.  \nonumber \\
&& \left. - 24\,{{{l_3^r}}^2} + 20\,{l_3^r}\,{l_4^r} +
 5\,{{{l_4^r}}^2} + {{6239}\over {331776\,{{\pi }^4}}} +  {1 \over {{\pi}^2}} \, \left(   -{{19}\over {3456}} - {{769} \over {576}}\, L
 - {{{1}}\over 6}\, l_1^r  \right. \right.
  \nonumber \\
 && \left. \left. +  {{{1}}\over 9}\, l_2^r - {{27} \over {16}}\,
 l_3^r  - {1 \over 8}\, l_4^r \right) + 3 r_1^r + 4 r_2^r + 4 r_3^r - 4
 r_4^r\right] \,{{M_{\pi}^4 \over F_{\pi}^4}}   \\
 & & \nonumber \\
\beta &=&1 + \left( -2\,L + 2\,{l_4^r} - {1\over {8\,{{\pi }^2}}} \right)
\,   {M_{\pi}^2 \over F_{\pi}^2} + \left[ {{5}\over 3}\, k_1 - {{5}\over 2}\, k_3 -
 3\,{k_4} - 4\,{l_3^r}\,{l_4^r} + 5\,{{{l_4^r}}^2}  \right.  \nonumber \\
&& \left. + {{8911}\over {331776\,{{\pi }^4}}} +   {1 \over {\pi}^2}\,
\left(  {-{1\over {512}}} + {{727}\over {864}}\, L  - {{11}\over
{18}}\, l_1^r -   {{7}\over 8}\, l_2^r - {{9}\over 8}\, l_3^r - { 1 \over
2}\, l_4^r \right)  \right.  \nonumber \\
&& \left. + r_2^r + 4 r_3^r -  4 r_4^r + 12 r_5^r - 4 r_6^r {{} \over {}}
\right] \,{{M_{\pi}^4 \over F_{\pi}^4}}   \\
&& \nonumber \\
\lambda_1 &=&-{1\over 3}\, L + 2\,{l_1^r} - {1\over {36\,{{\pi }^2}}} +
\left[ -{7\over 6}\, k_1 - {1\over 2}\, k_2 -      {1\over
3}\, k_4 + 8\,{l_1^r}\,{l_4^r} +  {{79}\over {9216\,{{\pi }^4}}}  \right.
  \nonumber \\
&& \left. + {1 \over 3456\, {\pi}^2}\, \left({{}\over{}}   -1 + 2272\,L - 2496\,{l_1^r} -
2160\,l_2^r - 384\,l_4^r \right) \right.  \nonumber \\
&& \left.  + r_3^r - r_4^r + 6 r_5^r - 2 r_6^r {{} \over {}}\right]
\,{M_{\pi}^2 \over F_{\pi}^2}  \\
&&  \nonumber \\
\lambda_2 &=&-{1\over 3}\, L  + {l_2^r} - {5\over {288\,{{\pi }^2}}} +
\left[ -{1\over 3}\, k_1 - {{4}\over 3}\, k_2 -     {1\over
3}\, k_4 + 4\,{l_2^r}\,{l_4^r} +      {{1223}\over {331776\,{{\pi }^4}}} 
\right. \nonumber \\
&& \left.    + {1 \over 27648\, {\pi}^2}\, \left( {{{}\over{}} 17 + 752\,L + 3840\,{l_1^r}
+     1536\,{l_2^r} - 1920\,{l_4^r}} \right) + 2 r_4^r\right] \,{M_{\pi}^2
\over F_{\pi}^2}   \\
&& \nonumber \\
\lambda_3 &=&-{{23}\over {18}}\, k_1 - {{37}\over {36}}\, k_2 +
{{277} \over {1990656\,{{\pi }^4}}} +  {1 \over 41472 \, {\pi}^2} \,
\left({{}\over{}} 19 + 5368\,L - 13056\, {l_1^r}  \right.  \nonumber \\
&& \left. - 9600\,{l_2^r} {{}\over{}} \right)  + r_5^r - {1 \over 3}\, r_6^r  \\
&&  \nonumber \\
\lambda_4 &=&{{5}\over {36}}\, k_1 + {{25}\over {72}}\, k_2 +
{{3311}\over {3981312\,{{\pi }^4}}} +   { 1 \over 10368 \, {\pi}^2}\, \left(
{{}\over{}} {-2 - 257\,L + 336\,{l_1^r} +
840\,{l_2^r}} \right) \nonumber \\
&& - {4 \over 3}\, r_6^r\ ,   
\end{eqnarray}
 with
\begin{equation}
{\mu}\,{{d\,l_i^r} \over {d\, \mu}} = - {{\gamma_i}
 \over {16 \pi^2 }}\ ,  \hspace{1cm} \gamma_1={1\over 3}\ ,  \gamma_2= {2
 \over 3}\ ,
 \gamma_3=-{1 \over 2}\ ,
  \gamma_4 = 2\ , 
  \end{equation}
and
\begin{equation}
L={{1}\over  {16 \pi^2}}\, \log{{{M_{\pi}^2} \over {\mu^2}}}\ ;  \hspace{1cm}   k_i(\mu)= (4
l_i^r (\mu ) - \gamma_i L) L\ .
\end{equation}
(These expressions are obtained from the expansions of the parameters $b_1$,
\ldots, $b_6$ originally given in \cite{bijn}, which are in one-to-one correspondence 
with $\alpha$, $\beta$, $\lambda_1$, \ldots, $\lambda_4$. We prefer to work  with
the latter set of parameters for the reader's convenience: explicit formulae for
low-energy observables in terms of $\alpha$, $\beta$, $\lambda_i$  are given
in Ref.~\cite{knecht}, whereas  similar expressions in terms of the 
$b_i$'s are at present not  available in the literature).
A few points are worth recalling. $i)$~The parameters $\alpha$, $\beta$,
$\lambda_i$ are $\mu$-independent. This fact,   together
with Eq.(10) fixes the scale dependence of the low-energy constants
$r_i^r(\mu)$. $ii)$~Eqs.~(4)-(9) fix the expansion of the
parameters $\alpha$, $\beta$, $\lambda_i$  in powers of
$M_{\pi}^2$ and $\log M_{\pi}^2$ (and/or in powers of the quark mass $m$),
since $l_1^r(\mu),\ldots,l_4^r(\mu)$ and the $r_i^r(\mu)$
are quark mass independent. Contributions of successive chiral orders to
$\alpha$, $\beta$, $\lambda_i$  can be identified  by counting the powers of
$M_{\pi}^2 / F_{\pi}^2$. Notice that $\alpha$ and $\beta$ start by an order
$O(p^2)$ contribution ($\alpha = 1, \beta = 1$) followed by $O(p^4)$ and
$O(p^6)$ corrections. The expansions of $\lambda_1, \lambda_2$ consist of
$O(p^4)$ and $O(p^6)$ contributions, whereas $\lambda_3$ and $\lambda_4$ are
entirely of order $O(p^6)$. $iii)$~The $O(p^4)$ constants $l_3$ and $l_4$ belong to
the explicit symmetry breaking sector of the effective lagrangian. They represent
 the fine tuning of the   $<\bar q q>$ condensate to its presumed
large value: in
S$\chi$PT, the deviation from the Gell-Mann--Oakes--Renner relation (3) is
given by \cite{gale}
\begin{equation}
{m \over m_0} = \left[ 2 l_3^r ( \mu) + 2 l_4^r(\mu) - {3 \over 2} L \right] \left( {M_{\pi} \over F_{\pi}}
\right)^2 + \ldots.
\end{equation}
Similarly, $l_4^r$ controls the deviation of $\beta$ from $1$. On the other
hand, the $\lambda_i$'s  are {\em independent} of $l_3^r$ (
and only  very weakly dependent on $l_4^r$) reflecting the fact that they
are only marginally  
sensitive  to the size of the $<\bar q q>$ condensate. In the sequel, we complete
our {\em definition of the standard $\chi$PT by adopting the
standardly used central values of $l_3^r$ and $l_4^r$} \cite{gale, bijn}:
\begin{equation}
l_3^r( M_{\rho}) = 0.82 \times 10^{-3}, \hspace{1cm} l_4^r (M_{\rho} ) = 5.6
\times 10^{-3}.
\end{equation}

Finally, the constants $l_1^r$ and $l_2^r$ do not describe
explicit symmetry breaking effects (they are coefficient of four-derivative terms in
${\cal{L}}^{(4)}$) and they are insensitive to the size of the quark condensate.
They control the parameters $\lambda_1$ and $\lambda_2$.

\vspace{.5cm}

{\bf 4}. Equations (4)-(9) can be used to predict the parameters $\alpha$,
$\beta$, $\lambda_1$, \ldots, $\lambda_4$  and consequently, all low-energy
$\pi \pi$
scattering observables, provided the  low-energy constants $l_1,
\ldots, l_4$ and $r_1, \ldots, r_6$  are determined from the analysis of
different processes. 
This is a path advocated by the authors of Ref.~\cite{bijn}.
In the present letter this kind of analysis will be confronted with additional
{\em experimental} information contained in Eq.~(2). Bijnens  {\it et al.} \cite{bijn} have used the values (13) for $l_3$ and $l_4$; for $l_1^r$ and $l_2^r$
they have taken the central values obtained from the S$\chi$PT analysis of $K_{l4}$
form factors \cite{kl4}:
\begin{equation}
l_1^r(M_{\rho}) = - 5.40 \times 10^{-3}, \hspace{1cm} l_2^r(M_{\rho}) =
5.67 \times 10^{-3}.
\end{equation}
As for the $O(p^6)$ constants $r_i^r(\mu)$, the authors of \cite{bijn} take
$r_i^r(1\mbox{GeV}) = 0$ and they check that this approximation confronted
with a resonance saturation model produces a negligible error. With the
values (14), and $r_i^r=0$ at $\mu=1\mbox{ GeV}$ one obtains (in this letter we always use $F_{\pi}=93.2 \mbox{ MeV}$ and $M_{\pi}=139.6 \mbox{ MeV}$):
\begin{eqnarray}
&\alpha = 1.074\ , & \beta = 1.105\ , \nonumber \\
& \lambda_1 = -8.91 \times 10^{-3}\ , & \lambda_2 = 14.5 \times 10^{-3}\ , \\
& \lambda_3 = 2.04 \times 10^{-4}\ , &
\lambda_4 = -1.79 \times 10^{-4}\ . \nonumber 
\end{eqnarray}

\centerline{\psfig{figure=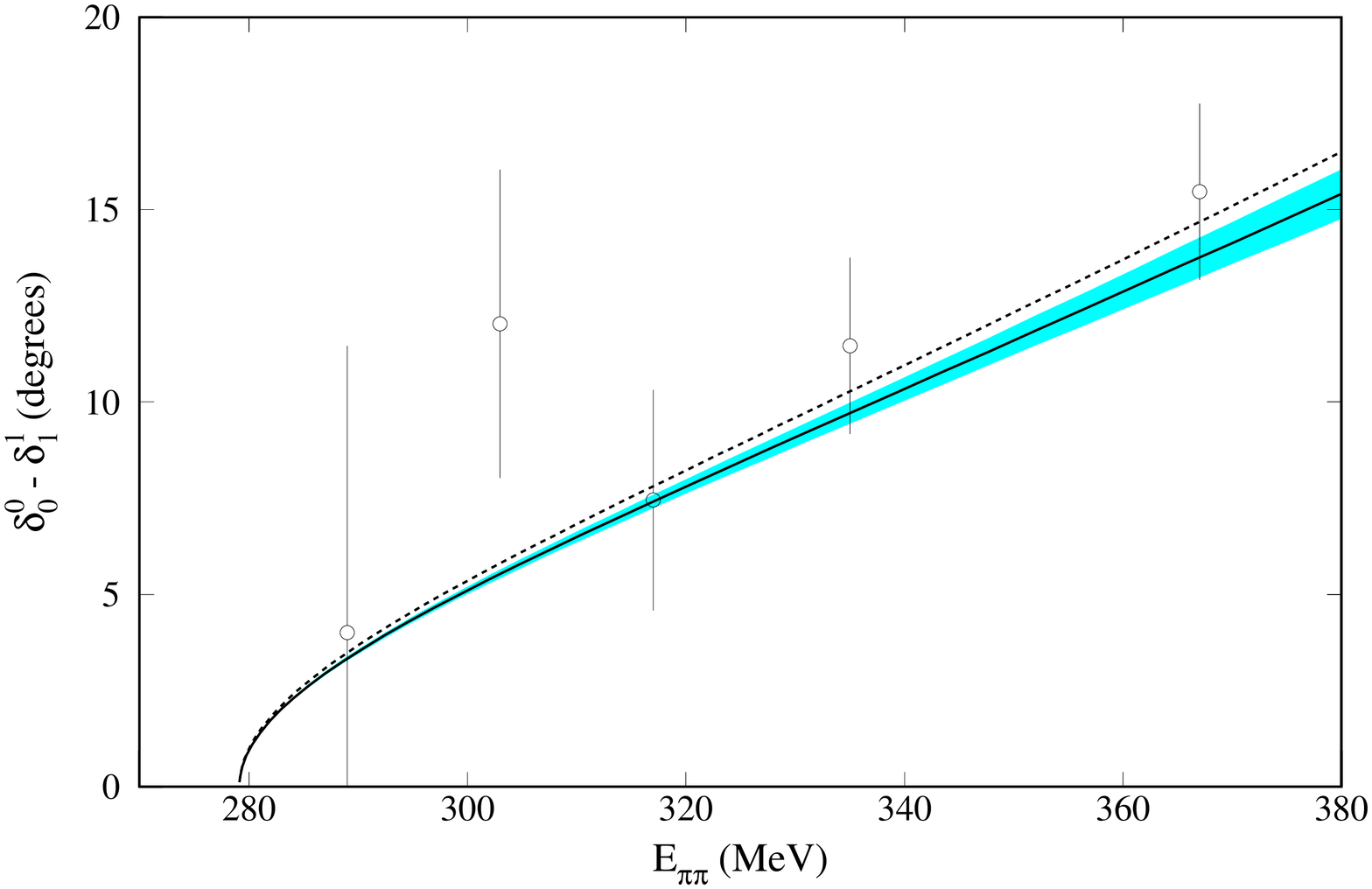,height=9.3cm,angle=-90}}
{\bf Figure 1}: {\it The phase shift difference $\delta_0^0 - \delta_1^1$ in the energy
region of $K_{l4}$ decays. The dashed curve is obtained with the values
of Eq.~(15) and it coincides with the curve displayed in Fig.~1 of
Ref.~\cite{bijn}.  The solid line is
obtained with the values of Eqs.~(2) and (18) while the shaded area results
adding the corresponding error bars quadratically. The experimental points
are  from Ref.~\cite{rosselet}.}

\indent

For these values of the parameters $\alpha$, $\beta$, $\lambda_1, \ldots,
\lambda_4$, one obtains the $S$-wave scattering lengths $a_0^0 = 0.218$,
$a_0^0 - a_0^2 = 0.259$ corresponding\footnote{Actually these have to be
compared with the numbers given in Eq.~(4) of Ref.~\cite{bijn} in
parentheses ($r_i^r(1 \mbox{ GeV})=0$). The small difference provides an
estimate of $O(p^8)$ effects: it is entirely due to the fact that the
amplitude $A_{KMSF}$ of Ref.~\cite{knecht} coincides with the amplitude
calculated in Ref.~\cite{bijn} only modulo $O(p^8)$ contributions.} to  the
predictions  given in Ref.~\cite{bijn}.
The resulting phase shift difference
$\delta_0^0  - \delta_1^1$
(measurable in $K_{l4}$ decays) is shown as a function of the center of mass
energy as the dashed line  in Fig.~1. It reproduces
 the  curve displayed in Fig.~1 of Ref.~\cite{bijn}. Finally, a few
remaining threshold parameters not discussed in Ref.~\cite{bijn} are
collected in the first column of Table~1, using the expressions displayed in
Appendix D of Ref.~\cite{knecht}.

\begin{center}
\begin{tabular}{|c||l|c|c|}
\hline
\hline
& Bijnens {\it et al.} \cite {bijn} &KMSF &Experiment \cite{experiment}\\
\hline
$a_0^0$ & $\ 0.218  \,(0.2156)$ & $0.209 \pm 0.004$ & $0.26 \pm 0.05$ \\
$ b_0^0$ & $\ 0.273  \,(0.271)$ & $0.255 \pm 0.010$ & $0.25 \pm 0.03$ \\
$ -10 a_0^2$ & \ $0.411\,  (0.4094)$ & $0.44 \pm 0.01$ & $0.28 \pm 0.12$ \\
$ -10 b_0^2$ & $\ 0.709 \,(0.704)$ & $0.80 \pm 0.02$ & $0.82 \pm 0.08$ \\
$a_0^0 - a_0^2$ & $\ 0.259\, (0.2565)$ & $0.254 \pm 0.004$ & $0.29 \pm 0.05$ \\
$10 a_1^1$ & $\ 0.395 \,(0.3956)$ & $0.373 \pm 0.008$ & $0.38 \pm 0.02$ \\
$ 10^2 b_1^1$ & $\ 0.785 \,(0.784)$ & $0.55 \pm 0.07$ & \\
$ 10^2 a_2^0$ & $\ 0.263 \,(0.267)$ & $0.16 \pm 0.02$ & $0.17 \pm 0.03$ \\
$ 10^3 a_2^2$ & $\ 0.237 \,(0.2356)$ & $0.09 \pm 0.13$ & $0.13 \pm 0.30$ \\
$10^4 a_3^1$ & $\ 0.428 \,(0.478)$ & $0.49 \pm 0.07$ & $0.6 \pm 0.2$ \\
\hline
\hline
\end{tabular}
\end{center}
{\bf Table~1}: {\it Threshold parameters of $\pi\pi$ scattering (in units of
$M_{\pi^+}$) in the standard
framework using the two-loop expressions of Ref.~\cite{knecht}, App. D. The
first column results from the values of Eq.~(15) (see the text for the
numbers in parentheses). The second column is obtained in the same way but
taking the values of Eqs.~(2) and (18) as input.}

\indent

The numbers in parentheses are obtained
keeping in higher orders only those components of $\alpha$, $\beta$,
$\lambda_1$ and $\lambda_2$ that actually contribute at most to the order
$O(p^6)$. These exactly coincide with the corresponding predictions one
would obtain using the amplitude given in \cite{bijn}. Among the latter it
is  worth
noticing the value predicted for the isoscalar $D$-wave scattering length
$a_2^0=26.3 \times 10^{-4}$, which is  three standard deviations above the
value extracted from the analysis of Roy equations \cite{experiment}. This
disagreement reflects the fact that the value (15) of $\lambda_2$ is
significantly above the value (2) inferred from experimental phase shifts in
Ref.~\cite{knecht2}. We would like to stress that both the canonical value
$a_2^0 = (17 \pm  3) \times 10^{-4}$ and the determination of the constant
$\lambda_2 = (9.3 \pm 0.5) \times 10^{-3}$ are based on the Roy dispersion
relations \cite{roy} using the experimental $\pi\pi$ data above $500\mbox{
MeV}$  as input.
Furthermore, in both cases, the dominant contribution comes from the
$P$-wave in the $\rho(770)$ region, which explains the relatively small
error bars. These facts suggest that the predictions of Ref.~\cite{bijn}
based on (15) have to be revised in order to agree with the values (2) of
the parameters $\lambda_1, \ldots, \lambda_4$ and with the standard value of
$a_2^0$.
We therefore proceed as follows: fixing $l_3^r$ and $l_4^r$ according to Eq.
(13), we solve Eqs.~(6) and (7) for $l_1^r(M_{\rho})$, $l_2^r(M_{\rho})$,
\begin{eqnarray}
 l_1^r(M_{\rho}) & = & (-4.0 \pm 1.0) \times 10^{-3} + \left[ -1.1 \, r_3^r + 1.0 \,
r_4^r - 6.3 \, r_5^r + 2.1 \, r_6^r {{}\over{}} 
\right]_{\mu=1\mbox{\scriptsize  GeV}}  \\
 l_2^r(M_{\rho}) & = & (1.6 \pm 0.4) \times 10^{-3} + \left[ {{}\over{}} 0.1 \, r_3^r - 3.5 \,
r_4^r +  0.5
\, r_5^r - 0.2 \, r_6^r \right]_{\mu=1\mbox{\scriptsize GeV}} 
\end{eqnarray}
where the values and errors (2) have been used for $\lambda_1$ and
$\lambda_2$. Eqs.~(16) and (17) are then inserted back into the formulae (4)
and (5) for $\alpha$ and $\beta$. Keeping in mind that $\alpha$ and $\beta$
are sensitive to $l_1$ and $l_2$ only at next-to-next-to-leading
level, the unknown constants $r_i^r (1\mbox{ GeV})$ are viewed as a source
of uncertainty in $\alpha$ and $\beta$. Inspired by na\"{\i}ve dimensional
analysis \cite{naive} we take in the expressions for $\alpha$ and $\beta$, $r_i^r ( 1 \mbox{ GeV})
= (0 \pm 2) \times 10^{-4}$. Adding the corresponding uncertainties quadratically,  we
obtain 
\begin{equation}
\alpha = 1.07 \pm 0.01 \hspace{1cm} \beta = 1.105 \pm 0.015.
\end{equation}
It should be stressed that the error in Eq.~(18) does not include the
uncertainty in the low-energy constants $l_3^r$ and $l_4^r$.
As in the case of the chiral condensate itself, the constant $l_3^r$ has not yet been
determined experimentally and for this reason it is hard to associate an
error bar with it. The values (18) have to be viewed as corresponding to the
``standard alternative'' of a large condensate {\em defined} by the values
(13) of $l_3^r$ and $l_4^r$. We now use the formulae given in Ref.~\cite{knecht} to generate the predictions for threshold parameters and phase
shifts that correspond to $\alpha$, $\beta$ (18) and $\lambda_1, \ldots,
\lambda_4$ (2). Adding the errors quadratically, the resulting threshold
parameters are summarized in the second column of Table~1. One observes that
the deviations of $a_0^0$ and $a_0^0 - a_0^2$ from their {\em central}
experimental values are significantly {\em larger} than   
predicted in Ref.~\cite{bijn}. Notice that now, the $D$-wave
scattering lengths perfectly agree with their Roy-equation ``experimental''
values as expected from the manner the values (2) of the constants $\lambda_1, \ldots,
\lambda_4$  have been obtained. 

\centerline{\psfig{figure=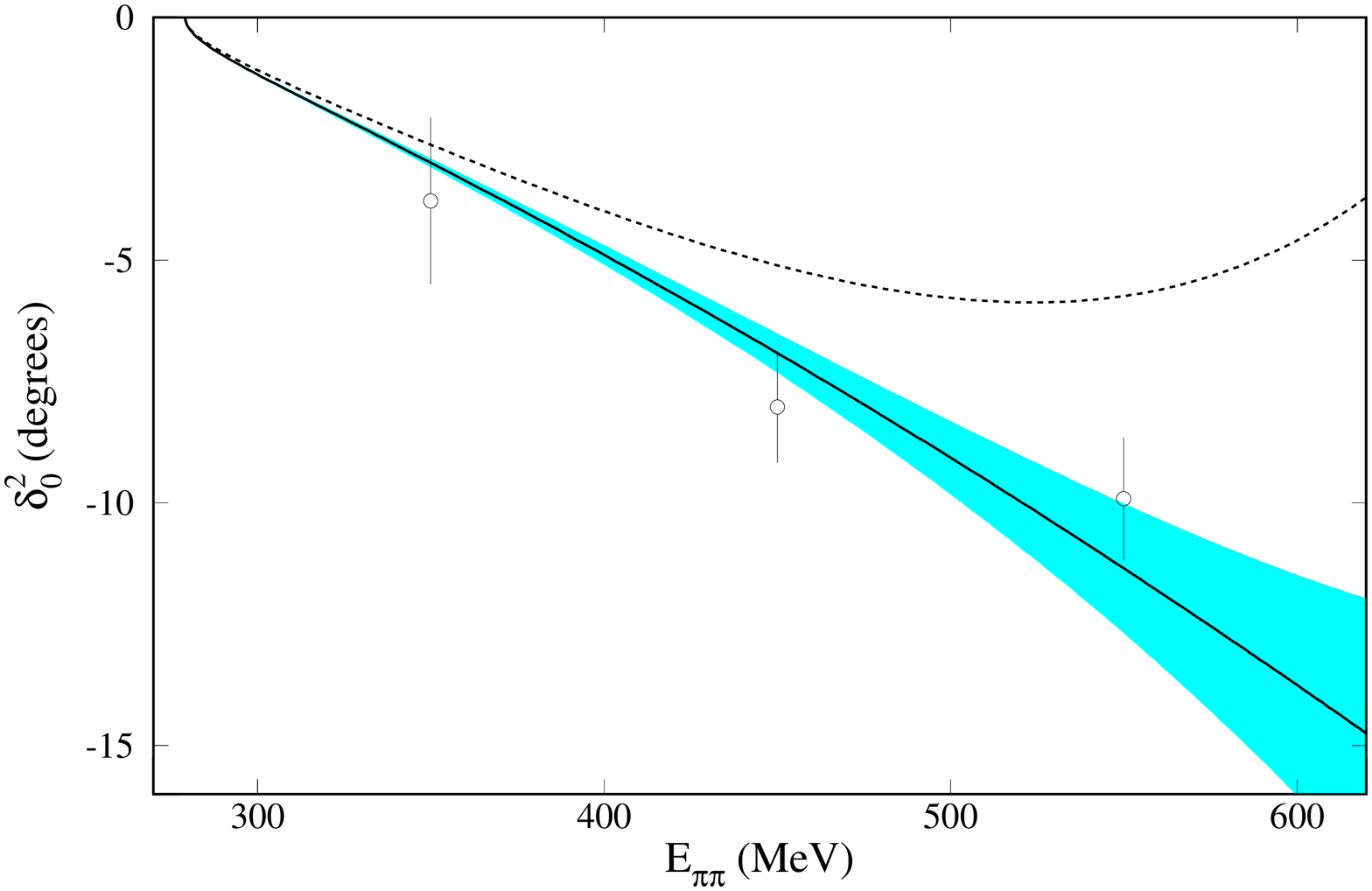,height=9.3cm,angle=-90}}
{\bf Figure 2}: {\it The isospin $2$, $S$-wave phase shifts at low energies. The
different curves are obtained with the values (15) (dashed) and the
values (2) and (18) (solid). In the latter case the shaded area shows the
corresponding error band. The experimental points are taken from
Ref.~\cite{hyams}.} 

\indent

A similar conclusion holds for the phase
shift difference $\delta_0^0 - \delta_1^1$, shown as the solid curve in
Fig.~1 with the error band indicated by the shaded area: the curve
displayed in Ref.~\cite{bijn} is significantly higher, i.e. closer to the
experimental central-value points. For illustration, the phase $\delta_0^2$
is also shown in Fig.~2.
 
\vspace{.5cm}

{\bf 5}. We finally address the question of interpreting the mismatch described in
the previous paragraph. Its origin clearly appears upon comparing eqs (16) and
(17) with the values of the constants $l^r_{1,2}(M_{\rho})$ extracted in
Ref.~\cite{kl4} from the ``unitarized'' one-loop S$\chi$PT $K_{l4}$ form
factors (Eq.~(5.10) of \cite{kl4}). Including errors the latter read:
\begin{equation}
l_1^r(M_{\rho}) = (-5.4 \pm 1.1) \times 10^{-3}, \hspace{1cm}
l_2^r(M_{\rho}) = (5.7 \pm 1.1) \times 10^{-3}.
\end{equation}
The question is how close   the expressions (16) and (17) can be brought to these
values keeping at the same time the $O(p^6)$ constants $r_3^r(1\mbox{ GeV}), \dots,
r_6^r(1\mbox{ GeV})$ at a reasonable size. If one proceeds as before
treating the $r^r_i$'s at 1~GeV as randomly distributed around zero with a standard
deviation $\pm 2 \times 10^{-4}$, one gets:
\begin{eqnarray}
l_1^r(M_{\rho}) &=& (-4.0 \pm 1.0 \pm 1.8) \times 10^{-3}\ ,   \nonumber \\
l_2^r(M_{\rho}) &=& (1.6 \pm 0.4 \pm 0.9) \times 10^{-3}\ ,
\end{eqnarray}
or 
\begin{equation}
{\bar l_1}= -0.37 \pm 0.95 \pm 1.71\ , \hspace{1cm} {\bar l_2} = 4.17 \pm 0.19
\pm 0.43\ ,
\end{equation}
where the first error has its origin in $\lambda_1$ and $\lambda_2$
(Eq.~(2)), whereas the second error arises from the presumed uncertainties in
the individual $r_i$'s added quadratically.
Two cheks of the size of the constants $r_i$ are conceivable.
$i)$ First, one can make a full use of informations contained in Eq.~(2)
determining the parameters $l_{1,2}^r(M_{\rho})$ and $r_3^r(1\mbox{ GeV}), \dots,
r_6^r(1\mbox{ GeV})$ by a simultaneous fit to Eqs.~(6)-(9) {\em
and} to the constraints $r_i^r (1 \mbox{ GeV}) = 0 \pm 2 \times 10^{-4}$. The
resulting $\chi^2 / d.o.f.$ is $1.9/2$ and one obtains 
\begin{equation}
l_1^r(M_{\rho}) =(-4.0 \pm 0.5) \times 10^{-3}, \hspace{1cm} l_2^r(M_{\rho})
= (2.0 \pm 0.3) \times 10^{-3},
\end{equation}
compatible with (20), whereas for the $r_i$'s one gets 
\begin{eqnarray}
&r_3^r(1\mbox{ GeV})=(-0.3 \pm 2.0) \times 10^{-4}\ , & r_4^r(1\mbox{
GeV})=(-0.7 \pm 0.9) \times 10^{-4}\ , \nonumber \\ 
&& \\
&r_5^r(1\mbox{ GeV})=(1.5 \pm 0.5) \times 10^{-4}\ , & r_6^r(1\mbox{ GeV})=(0.4
\pm 0.9) \times 10^{-4}\ . \nonumber 
\end{eqnarray}
This result turns out to be rather stable: if one increases the
uncertainties in the $r_i$'s to $\pm 3 \times 10^{-4}$, the new $\chi^2 /
d.o.f. = 1.24 / 2$, the values (22) become $(-4.8 \pm 0.5) \times 10^{-3}$
and $(2.1 \pm 0.3) \times 10^{-3}$
respectively, and the changes in the $r_i$'s also remain rather modest. On the
other hand, the errors obtained by this procedure (increase of $\chi^2$ by
one unit) and shown in Eqs.~(22) and (23) are probably heavily underestimated.
$ii)$ Next, it is instructive to confront the previous discussion with the
estimate of the constants $r_i$ by resonance saturation as quoted recently
by Hannah \cite{hannah}:
\begin{eqnarray}
r_1=-0.61 \times 10^{-4}, & r_2 = 1.3 \times 10^{-4}, \nonumber \\
r_3=-1.70 \times 10^{-4}, & r_4 = -1.0 \times 10^{-4}, \\
r_5=1.14 \times 10^{-4}, & r_6 = 0.3 \times 10^{-4}. \nonumber
\end{eqnarray}
Estimating low-energy constants by resonance saturation does not, in
principle, fix the renormalization scale $\mu$ at which the estimate is
supposed to hold. Actually, if a constant exhibits a strong scale
dependence, its resonance saturation estimate is subject to caution.
Interpreting Eqs.~(24) as values of $r_i^r(\mu)$ at $\mu = 1 \mbox{ GeV}$,
one observes a striking coherence with the preceding analysis: (24) is,
indeed, consistent not only with dimensional analysis or with the assumption
$|r_i^r| < 2\times 10^{-4}$ but, moreover it agrees with the fit (23). One
can even repeat the fit to Eqs.~(6)-(9) constraining
$r_i^r(1\mbox{ GeV})$ to the values (24) allowing for a $100\%$ error: the fit
is excellent ($\chi^2 / d.o.f. = 0.91/2$) and it yields $l_1^r(M_{\rho}) =
(-4.0 \pm 0.5) \times 10^{-3}$, $l_2^r(M_{\rho}) = (2.1 \pm 0.3) \times
10^{-3}$, again compatible with (20) and (22). On the other hand, one finds
that between $\mu = M_{\rho}$ and $\mu=1$ GeV, only the constants $r_4$,
$r_5$ and $r_6$ show a moderate scale dependence: had we assumed that the
values (24) concern the scale $\mu = M_{\rho}$ (as suggested in Ref.~\cite{hannah}), the comparison with our previous analysis would be less
favourable as far as the constant $r_3$ is concerned, $r_3^r(1 \mbox{ GeV}) =
-4.9 \times 10^{-4}$ in this case. Notice however   that according to Eq.~(17)
the correction to the ``critical'' constant $l_2^r(M_{\rho})$ is dominated by
$r_4^r$ whose scale dependence is rather weak:
\begin{equation}
r_4^r(1 \mbox{ GeV}) = r_4^r(M_{\rho}) -7 \times 10^{-6}.
\end{equation}
In order that the constant $l_2^r(M_{\rho})$ (17) differ from  the $K_{l4}$
value (19) by at most two  standard deviations, the constant $r_4^r(1\mbox{
GeV})$ would have to be $r_4^r(1 \mbox{ GeV}) \simeq -5 \times 10^{-4}$.
This cannot be excluded but it looks unlikely in the light of the present
analysis.

\vspace{.5cm}

{\bf 6}. The constants $l_1$ and $l_2$ (19) have not been obtained from a full
two-loop analysis of $K_{l4}$ form factors $F$ and $G$, which is not yet
available. Instead, their determination is based on matching a dispersive
representation for the form factor $F$ with the one-loop S$\chi$PT
expressions, the latter merely serving to fix the subtraction constants.
This method of ``improving'' one-loop $\chi$PT calculations has been often
used in the past \cite{matching} and it suffers from a basic ambiguity: one has to assume
that the one-loop and  two-loop amplitudes practically coincide in a
particular kinematical point $M$. Even if one admits the very existence of
such a matching point $M$, the results can still depend on its choice. In
Ref.~\cite{kl4} the matching point has been chosen at the threshold $s_{\pi}= 4
M_{\pi}^2$  of the $S$-wave amplitude $\pi\pi \rightarrow K + \mbox{axial
current}$, where $s_{\pi}$ stands for the dipion invariant mass squared. We
have repeated the analysis of Ref.~\cite{kl4} for other choices of the
matching point between $s_{\pi}= 4 M_{\pi}^2$
and the left-hand-cut branch point $s_{\pi}=0$. We reproduce the result
(19) and find that it is actually rather insensitive to the matching point
except in the vicinity of the singular point $s_{\pi}=0$, where the outcome
for $l_1$ (but not $l_2$) becomes less stable.
For instance, with the matching point at $s_{\pi}= 2 M_{\pi}^2$, we obtain
\begin{equation}
l_1^r(M_{\rho}) = (-4.8 \pm 2.1) \times 10^{-3}, \hspace{1cm}
l_2^r(M_{\rho}) = (5.3 \pm 1.0) \times 10^{-3}.
\end{equation}
Given the present state and quality of $K_{l4}$ experimental data, it seems
hard to ascribe the  discrepancy described above to the S$\chi$PT
analysis performed in Ref.~\cite{kl4}. On the other hand, it should be kept
in mind that outside the standard framework, i.e. for low values of the
condensate $<\bar q q >$, the constants $l_1$ and $l_2$ extracted from
$K_{l4}$ data will be modified already at the one-loop level: since in
G$\chi$PT the loop contributions are more important, the resulting central values of
$|l_1|$ and $|l_2|$ are expected to come out somewhat smaller \cite{workshop}.

\vspace{.5cm}

{\bf 7}. A few concluding remarks are in order. The past determinations
\cite{gale,egpr,ananth,pp} of the constants $l_1^r$ and $l_2^r$ have operated
within the $O(p^4)$ order of $\chi$PT. They have shown an apparent coherence
and compatibility with the $K_{l4}$ analyses of Ref.~\cite{kl4}. This
compatibility might be lost at $O(p^6)$ order and we have to understand why.
The resonance saturation models are the only ones that determine the
constants $l_{1,2}$ directly, integrating out the resonance degrees of
freedom from an extended effective lagrangian ${\cal L}_{eff}$. However, incorporating resonances
into ${\cal L}_{eff}$ is not free of ambiguities, especially if one aims
at the $O(p^6)$ accuracy. On the other hand, less model-dependent sources of
information, such as $\pi\pi$ $D$-waves \cite{gale} and/or sum rules
\cite{ananth,pp} primarily determine the physical parameters $\lambda_1,
\lambda_2$. It turns out that this determination is rather stable and barely
affected by switching from order $O(p^4)$ to $O(p^6)$. At the $O(p^4)$ level,
 i.e. neglecting in Eq.~(1) the two-loop effects and setting $\lambda_3 =
\lambda_4=0$, one would get from the $a_2^0$ and $a_2^2$ experimental central
values $\lambda_1 = -6.4 \times 10^{-3}$ and  $\lambda_2= 10.8 \times
10^{-3}$,
to be compared with Eq.~(2). In other words, the relationship between
$D$-wave scattering lengths and the parameters $\lambda_1, \lambda_2$ is almost
unaffected by $O(p^6)$ effects. The latter however become rather important in
the relationship between $\lambda_2$ and $l_2^r$. Rewriting Eq.~(7) to make
the dependence on $l_2^r(M_{\rho})$  appear explicitly, one obtains
\begin{equation}
\lambda_2= \{ l_2^r(M_{\rho}) + 5.45 \times 10^{-3} \} + \{
0.32 \times l_2^r(M_{\rho}) + 1.7 \times 10^{-3} \}
\end{equation}
where the first (second) curly brackets collect all $O(p^4)$ ($O(p^6)$)
contributions ($r_4$ has been neglected). The $O(p^6)$ contribution is as
large as $30 \%$ and it is dominated by double logs, whose importance has
been anticipated by Colangelo \cite{colangelo}. It follows that for a given
$\lambda_2$ ($D$-waves), the resulting value of $l_2^r(M_{\rho}) $ can easily differ
by a factor $\sim 2$ depending whether in Eq.~(27) one includes the $O(p^6)$
term or not. Whether the consistency with $K_{l4}$ form factors can be
understood within the large condensate hypothesis remains to be clarified.
It might be, for instance, that at $O(p^6)$ level the $K_{l4}$ form factors
also 
receive an important contribution from double logs, which the unitarization
procedure would not take into account \cite{leut}.
Independently of this issue, the main conclusion of this letter is the
following: the predictions of S$\chi$PT for $a_0^0$, $a_0^0 -a_0^2$ and
$\delta_0^0 - \delta_1^1$ given in Ref.~\cite{bijn} are systematically
overestimated as shown in Fig.~1 and Table~1 of the present paper. A closely
related fact is the failure of the values of $l_1^r$ and $l_2^r$ used in
Ref.~\cite{bijn} to describe the $D$-waves in agreement with Roy  equations
analyses. This agreement is nicely recovered if instead the present
determinations of Eq.~(20) are used. 
This shows, once more, that a sensible and sensitive test of QCD in low-energy
$\pi\pi$ scattering should be based on a global analysis making use of all
theoretical constraints and all pertinent low-energy observables.

\clearpage
\newcommand{\np}[1]{Nucl.\ Phys.\ {\bf #1}}
\newcommand{\prep}[1]{Phys.\ Rep.\ {\bf #1}}
\newcommand{\pr}[1]{Phys.\ Rev.\ {\bf #1}}
\newcommand{\prl}[1]{Phys.\ Rev.\ Lett.\ {\bf #1}}
\newcommand{\pl}[1]{Phys.\ Lett.\ {\bf #1}}
\newcommand{\ap}[1]{Ann. Phys. (NY)\ {\bf #1}}

\newcommand{\RMP}[1]{Rev.\ of Mod.\ Phys.\ {\bf #1}}
\newcommand{\NC}[1]{Nuovo Cimento {\bf #1}}
\newcommand{\CMP}[1]{Comm.\ Math.\ Phys.\ {\bf #1}}
\newcommand{\MPL}[1]{Mod.\ Phys.\ Lett.\ {\bf #1}}
\newcommand{\BLMS}[1]{Bull.\ London Math.\ Soc.\ {\bf #1}}


\end{document}